\documentclass[prl,showpacs,groupedaddress,superscriptaddress,twocolumn]{revtex4}

\usepackage{fontenc,times,amsmath,amsfonts,amssymb,graphics,graphicx,epsfig,color,bbm,natbib,bm}

\begin{document}

\title{Classical Fisher Information in Quantum Metrology -- Interplay of Probe, Dynamics and Measurement}

\author{Gabriel A.\ Durkin} \email{gabriel.durkin@qubit.org}
\affiliation{Quantum Laboratory , NASA Ames Research Center,  Moffett Field, California 94035, USA}

\date{\today}

 
\begin{abstract}  We introduce a positive Hermitian operator, the Fisher operator, and use it to examine a measurement process incorporating unitary dynamics and complete measurements. We develop the idea of information complement, the minimization of which establishes the optimal precision for a fixed input.  The formalism demonstrates that, in general, the classical Fisher Information has the Hamiltonian semi-norm as an upper bound. This is achievable with a qubit probe and only projective measurements, and is independent of the true value of the estimated parameter. In an interferometry context, we show that an optimal measurement scheme can be constructed from linear optics and photon counting, without recourse to generalised measurements or exotic unitaries outside of $SU(2)$.  \end{abstract}

\maketitle

Scientists strive  for an understanding of Nature by a physical interaction introducing correlations between observer and observed, and the process is called measurement. Quantum mechanics exerts fundamental limitations on the precision of any measurement \cite{Luis}, yet for a quantum observable there is still a classical probability distribution over measurement outcomes. This distribution may depend on some real-valued system parameter $\theta$ such as an interaction time or phase angle. It may have no associated Hermitian observable, nor be measurable directly -- but  its estimation may be the true goal of the measurement. Inferring $\theta$ from frequencies of measurement outcomes is a conventional challenge in classical information theory with a well-established methodology \cite{Cover and Thomas}. From this classical underpinning we will show a quantum formalism emerges directly and without recourse to the quantum Fisher Information \cite{Metrology,MetrologyII} or a linear error propagation model \cite{HLimited-Interfer-Decorr,Yurke-SU2-Interfer}. Those approaches focus on the geometry of the state in Hilbert space and have a long distinguished history stretching at least as far back as 1976 \cite{Helstrom}. In contrast, the formulation in this paper is self-contained and incorporates all three instrument components --  input, dynamics and measurement choice -- on an equal footing. Our conclusions rely on no other results or assumptions than those presented herein.

Consider a maximal test  \cite{Peres} having outcomes labelled $k$ and an associated probability distribution $P(\theta)=\{p_k (\theta)\}$  that depends on a continuous real parameter $\theta$. Classical Fisher information \cite{Cramer-Fisher} is defined as
\begin{equation}
\mathcal{J}(\theta) = \sum_{k} p_k (\theta)\left(  \frac{\partial }{\partial \theta} \ln p_k (\theta) \right)^2 \; . \label{F1}
\end{equation}
This  functional is a measure of the information contained in the distribution $P(\theta)$ about the parameter $\theta$ \cite{Cover and Thomas}. It provides the sole distance metric in probability space \cite{Fisher-unique-metric} and the scaling factor in local distinguishability \cite{Durkin-Dowling}. An explicit lower bound on the standard error of an unbiased estimate $\tilde{\theta}$ on the true phase $\theta$ is given by the reciprocal of the Fisher information, $(\delta \tilde{\theta})^2 \geq 1 / \mathcal{J}(\theta) $,  called the Cram\'{e}r-Rao bound \cite{Cramer-Fisher}. Therefore, for optimal precision one maximizes the Fisher information \cite{Braunstein-knee}.

Let's take the case of the measurement observable $\hat{M}$ and outcomes $\{m_k\}$ associated with distribution $\{p_k\}$ applied to a quantum system that previously evolved from a known initial state $|\psi_{0} \rangle$ under the dynamics of some Hamiltonian $\hat{H}$ for time $\theta$.  The Schr\"{o}dinger equation governs the dynamics $ i \partial_{\theta}  | \psi_{\theta} \rangle = \hat{H} | \psi_{\theta} \rangle $,
(where $\hbar=1$) and the time evolution is explicitly $ | \psi_{\theta} \rangle = \exp\{- i \hat{H} \theta\} | \psi_{0} \rangle \; .$

The parameter estimation task involves an inference of the time-like variable $\theta$ from the measurement distribution $\{ p_k (\theta)\}$. Writing the spectral decomposition of the maximal measurement as $ \hat{M} =  \sum_{k} m_k | k \rangle \langle k | $
then  complex amplitudes $\langle k | \psi_{\theta} \rangle = r_k \exp\{i \phi_k\}$
lead to probabilities $ p_k = \langle k | \psi_{\theta} \rangle \langle \psi_{\theta} |  k \rangle = r_{k}^{2} \; , \label{probs}$ where it is understood that  $\{p_k, r_k \} \in [0,1]$ and $\phi_k \in [0,2 \pi )$ are all real-valued functions of parameter $\theta$. Replacing $p_k$ with $r_k^2$ in Eq.\eqref{F1} gives:
\begin{equation}
\mathcal{J}(\theta)= 4\sum_k \dot{r}_{k}^{2}  \; ,\label{FI2}
\end{equation}
which we can now use to find an operator expression for the Fisher information. Differentiating $r_k^2$ gives
\begin{equation}
2 \dot{r}_k r_{k}= \langle k | \psi_{\theta} \rangle \langle  \dot{\psi}_{\theta} |  k \rangle + \langle k | \dot{\psi}_{\theta} \rangle \langle \psi_{\theta} |  k \rangle  \; . \label{diff}
\end{equation}
Now, $ \langle k | \dot{\psi}_{\theta} \rangle  = \partial_{\theta} (r_k e^{i \phi_k})= e^{i \phi_k} (\dot{r}+i r \dot{\phi}_k) =  | \langle k | \dot{\psi}_{\theta} \rangle| e^{i ( \phi_k + \tau_k)} $ where we define the inclination of the velocity vector:
\begin{equation}
\tau_k = \arg  \langle k | \dot{\psi}_{\theta} \rangle  - \arg \langle k | \psi_{\theta} \rangle = \tan^{-1}( r_k \dot{\phi}_k / \dot{r}_k) \; .
\end{equation}
Eq. \eqref{diff} yields  $\dot{r}_k  = \cos \tau_k  | \langle k | \dot{\psi}_{\theta} \rangle|  $. From the Schr\"{o}dinger equation we also have $ i \langle k | \dot{\psi}_{\theta} \rangle =  \langle k | \hat{H} |  \psi_{\theta} \rangle $. Substituting this expression and squaring gives
\begin{equation}
\dot{r}_{k}^{2}  = \cos^2 \tau_k    \langle  \psi_{\theta} | \hat{H} | k  \rangle  \langle  k | \hat{H} | \psi_{\theta} \rangle   . 
\end{equation}
 Summing over all the measurement outcomes `$k$' gives the Fisher information using Eq.\eqref{FI2}: $\mathcal{J}(\theta) = 4 \sum_k   \cos^2 \tau_k   \langle  \psi_{\theta} | \hat{H} | k  \rangle   
  \langle  k | \hat{H} | \psi_{\theta} \rangle $. Therefore, we can define a non-linear, positive and hermitian Fisher operator $\hat{F}_{\theta}$ that is diagonal in the measurement basis:
\begin{equation}
\hat{F}_{\theta} = 4 \sum_k \cos^2  \tau_k | k \rangle \langle k | = 4 \sum_k c_{\psi,k}  | k \rangle \langle k |   \: , \label{FOp}
\end{equation}
such that the Fisher information is then
\begin{equation}
\mathcal{J}(\theta) =   \langle  \psi_{\theta} | \hat{H} \hat{F}_{\theta}  \hat{H} | \psi_{\theta} \rangle = \text{Tr} \; \left[ | \psi_{\theta} \rangle  \langle  \psi_{\theta} |  \: \hat{H} \hat{F}_{\theta}  \hat{H} \right ] . \label{FisherInfo}
\end{equation}
This can be re-expressed as an expectation value with respect to the probe state, the input $| \psi_0 \rangle$, by unitarily transforming the Fisher operator:
$
\mathcal{J}(\theta) =   \langle  \psi_{0} | \hat{H} \hat{\Phi}_{\theta}  \hat{H} | \psi_{0} \rangle =  \text{Tr}  \; \left [ | \psi_{0} \rangle  \langle  \psi_{0}   | \:  \hat{H} \hat{\Phi}_{\theta}  \hat{H} \right ] ,
$
where
\begin{equation}
\hat{\Phi}_{\theta} = e^{i \hat{H} \theta}   \hat{F}_{\theta} e^{- i \hat{H} \theta}  =  4 \sum_k c_{\psi,k}  | k'  \rangle \langle k' | 
\end{equation}
is the unitarily transformed Fisher operator, cf Eq. \eqref{FOp}. Note that, due to the non-linear nature of $\hat{F}_\theta$, a basis transformation $| k \rangle \mapsto | k' \rangle$ would give a different result: 
\begin{equation}
\hat{F}_\theta \mapsto \hat{F}_\theta' = 4 \sum_k c_{\psi,k'}  | k'  \rangle \langle k' | \; ,
\end{equation} 
not equivalent to $\Phi_\theta$, since $c_{\psi,k'}  \neq c_{\psi,k} $ generally.

\emph{Fixed Probe Optimization}: The completeness of the measurement basis provides a resolution of the identity  $ \sum_k | k \rangle \langle k | = \mathbbm{1} $,  and therefore Eq.\eqref{FisherInfo} becomes:
\begin{align}
\frac{\mathcal{J}(\theta) }{4} & = \langle \hat{H}^2 \rangle - \sum_k \sin^2  \tau_k \:  \langle  \psi_{\theta} | \hat{H} | k  \rangle   
  \langle  k | \hat{H} | \psi_{\theta} \rangle  \nonumber \\
   & =  \langle \hat{H}^2 \rangle - \sum_k ( r_k \dot{\phi}_k )^2 =  \langle \hat{H}^2 \rangle - \mathcal{K}(\theta) \; ,    \label{Fisher1}
\end{align} 
where the form in the second line was previously presented in \cite{Durkin-Dowling}. With reference to Eq.\eqref{FI2} it is interesting to note that $\langle \hat{H}^2 \rangle$ is a sum of translational and rotational kinetic energy terms. Here it may be taken with respect to $| \psi_0 \rangle$, since $[\hat{H}^2, \exp\{i \hat{H} \theta \}] = 0$. In Eq.\eqref{Fisher1} the subtracted term, we will call it the `\emph{information complement}' $\mathcal{K}$, is positive and can only reduce $\mathcal{J}(\theta)$. We look for a measurement basis $\{ | k \rangle \}$ and associated set $\{ r_k, \phi_k\}$ that minimizes this information complement for a fixed input $|\psi_0\rangle$.  Finding derivatives of $\phi_k$ by writing $\delta \phi_k = \arg \langle k | \psi_{\theta + \delta \theta}  \rangle - \arg \langle k | \psi_{\theta}   \rangle$ produces
\begin{equation}
r_k \dot{\phi}_k = \sum_l | H_{k , l} | r_l \cos (\phi_l - \phi_k + \Omega_{k,l}) = \mathcal{A}_k \; ,
\end{equation}
where $\langle  k| \hat{H} | l \rangle =  H_{k , l}  = | H_{k , l} |  \exp i  \Omega_{k,l}$, and therefore
\begin{equation}
\mathcal{K} =  \sum_k ( r_k \dot{\phi}_k )^2  = \sum_k \mathcal{A}_k^2 \; .
\end{equation}
By comparison, expanding $\langle \hat{H} \rangle$ in the $| k \rangle$ basis gives
\begin{equation}
\langle \hat{H} \rangle = \sum_{k,l} r_l r_k  | H_{k , l} | r_l \cos (\phi_l - \phi_k + \Omega_{k,l})  = \sum_k r_k \mathcal{A}_k
\end{equation}
At the minimum
\begin{equation}
\frac{\partial  \mathcal{K} }{\partial \phi_p}= \sum_k \frac{\partial}{\partial \phi_p}( \mathcal{A}_k^2) \;  = \; 0
\end{equation}
and this calculation leads to
\begin{equation}
\mathcal{A}_p = - r_p \frac{\sum_k \mathcal{A}_k  | H_{k , p} |  \sin (\phi_p - \phi_k + \Omega_{k,p})  }{\sum_k r_k  | H_{k , p} |  \sin (\phi_p - \phi_k + \Omega_{k,p}) } = r_p \mathcal{B}_p \; ,
\end{equation}
at the minimum of $\mathcal{K}$. Remember, this is equivalent to maximizing $\mathcal{J}$ over $\{ | k \rangle \}$ for a fixed $| \psi_0 \rangle$. At this minimum we can prove the relation that
\begin{align}
 &\mathcal{K} - \langle \hat{H} \rangle^2 \nonumber \\
= & \sum_k \mathcal{A}_k^2 -  ( \sum_k r_k \mathcal{A}_k )^2  \nonumber \\
= & \sum_k  r_k^2 \mathcal{B}_k^2 -  ( \sum_k r_k^2 \mathcal{B}_k )^2  \nonumber \\
= &  \: \Delta^2 \mathcal{B} \geq 0 \; ,
\end{align} since $\{r_k^2\} =\{ p_k\}$ is a probability distribution. Comparing again with Eq.\eqref{Fisher1} it follows directly that
\begin{equation}
\mathcal{J}(\theta) /4 \leq \Delta^2 \hat{H}
\end{equation}
for a fixed input  $| \psi_0 \rangle$. Now we will show that this bound is saturated by a particular qubit input state.

\begin{figure}
\includegraphics[width=2.8in]{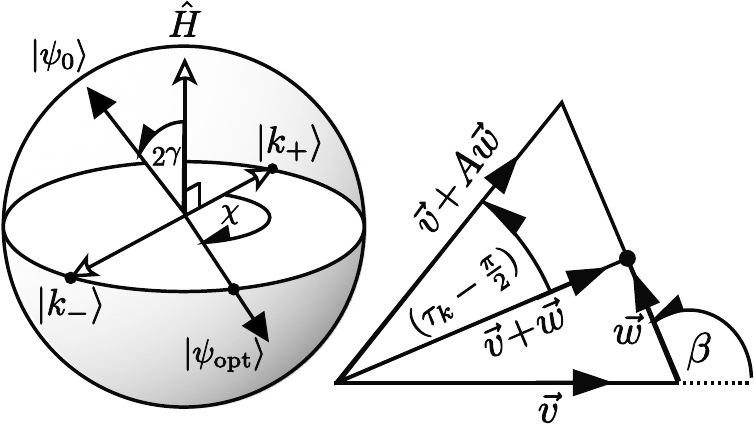}
\caption{\small{ \emph{Left}: The optimal measurement scheme for any Hamiltonian $\hat{H}$ may be restricted to the qubit subspace of its extremal eigenvectors $| \lambda_{\uparrow}\rangle$ and $| \lambda_{\downarrow} \rangle$. On the Bloch sphere: the diagonal basis of $\hat{H}$ defines the $z$-axis, and the optimal measurement projectors $| k_{\pm} \rangle$ the $x$-axis. An optimal probe state $| \psi_{\text{opt}}\rangle$ lies anywhere in the equatorial plane.  \emph{Right}: The vector construction for finding $c_{\psi,k}$ in Eq.\eqref{coeffs} with $A>1$.}} \label{sphere}
\end{figure}

\emph{Optimizing for a Qubit:}  For probe states that are eigenstate of the Hamiltonian  then all  $\dot{r}_k \mapsto 0 $ and $\mathcal{J} = 0 $ from Eq.\eqref{FI2} because any eigenstate $| \lambda \rangle$ of $\hat{H}$ only gains a phase during its evolution. Thus $r_k (\theta)= |\langle k | e^{i \hat{H} \theta}| \lambda \rangle | =  |\langle k | e^{i \lambda \theta}| \lambda \rangle | =  |\langle k | \lambda \rangle | = r_k(0)$, and $\dot{r}_k = 0$.
For optimality over all $| \psi_0 \rangle$ and $\{ | k \rangle \}$ the input state must thus be a \emph{superposition} of at least two Hamiltonian eigenvectors, 
 \begin{equation}
 | \psi_0 \rangle \mapsto \cos \gamma  |  \lambda_{1} \rangle + e^{i \chi} \sin \gamma | \lambda_2 \rangle  \; .  \label{general-probe}
 \end{equation}
Starting with a qubit probe state of the form of Eq. \eqref{general-probe} and some measurement basis pair $\{ | k_1 \rangle, | k_2 \rangle\}$ spanning the same  $\mathbbm{C}^2$ as $\{ | \lambda_1 \rangle, | \lambda_2 \rangle\}$:
\begin{align}
| k_1 \rangle = & \: \; \; \; \cos \alpha | \lambda_1 \rangle  + \sin  \alpha | \lambda_2 \rangle \nonumber \\
| k_2 \rangle =  & -  \sin \alpha | \lambda_1 \rangle + \cos \alpha | \lambda_2 \rangle
\end{align}
Here it has been chosen that the measurement basis defines the $x$ axis on the Bloch sphere of FIG.\ref{sphere}, hence the real coefficients for $| k_{1,2}\rangle$. By confining the measurement basis  vectors $\{ | k_1, \rangle, | k_2 \rangle \}$ to the same space as  $\{ | \lambda_1 \rangle, | \lambda_2 \rangle\}$ then one can restrict interest to the component of   $\hat{\Phi}_\theta$ within this qubit space:
\begin{align}
\hat{\Phi}_\theta =  & \; e^{i \hat{H} \theta} \{  c_{\psi,k1}  | k_1 \rangle \langle k_1  | +  c_{\psi,k2} | k_2 \rangle \langle k_2  | + \dots \nonumber \} e^{- i \hat{H} \theta}  \nonumber \\
=  & \; \;   c_{\psi,k1}  \left(
\begin{array}{ll}
c^2  & e^{-i ( \lambda_2-\lambda_1) \theta } s c \\
e^{+i  (\lambda_2-\lambda_1) \theta } s c & s^2
\end{array}
\right) \nonumber \\ + & \;  c_{\psi,k2}  \left(
\begin{array}{ll}
  s^2  &  - e^{-i  (\lambda_2-\lambda_1) \theta) } s c \\
- e^{+ i  (\lambda_2-\lambda_1) \theta)}  s c & c^2 
\end{array}
\right) +  \dots \nonumber
\end{align}
where $ c $  ($s$) is $ \cos \alpha$  ($ \sin \alpha$).  We ignore elements of $\hat{\Phi}_\theta$ that project onto the remaining Hilbert space, orthogonal to   $\{ | \lambda_1 \rangle, | \lambda_2 \rangle\}$.
The coefficients $\cos^2 \tau_k$ have a geometric interpretation by mapping $\mathbbm{C} \mapsto \mathbbm{R}^2$:
\begin{align} 
 \tau_k &= \cos^{-1}\frac{\langle k | \dot{\psi}_\theta \rangle . \langle k | \psi_\theta \rangle}{|\langle k | \dot{\psi}_\theta \rangle | | \langle k | \psi_\theta \rangle|} = \sin^{-1} \frac{\langle k | \hat{H} | \psi_\theta \rangle . \langle k | \psi_\theta \rangle}{|\langle k | \hat{H} | \psi_\theta \rangle | | \langle k | \psi_\theta \rangle|} \nonumber \\  &=  \sin^{-1}\frac{(\vec{v} + A \vec{w}) . (\vec{v} + \vec{w}) }{ | \vec{v} + A \vec{w} |  |\vec{v} + \vec{w} |} \: ,  \end{align}
with $A = \lambda_2 / \lambda_1$ and
\begin{align}
\vec{v}_1 & =  \cos( \alpha) \cos ( \gamma) , \; \; \vec{w}_1 =  \sin (\alpha) \sin (\gamma) e^{  i (\chi - (\lambda_2 - \lambda_1) \theta ) } \nonumber \\
\vec{v}_2 & =   - \sin ( \alpha) \cos ( \gamma), \; \vec{w}_2 =   \cos (\alpha) \sin (\gamma) e^{  i (\chi  - (\lambda_2 - \lambda_1) \theta ) } \ \nonumber .
\end{align}
The components in the measurement basis are:
\begin{align}
\langle k_{1,2}| e^{-i \hat{H} \theta} | \psi_{0} \rangle = e^{-i \lambda_1 \theta} (\vec{v}_{1,2} +  \vec{w}_{1,2}  ) \nonumber \\
\langle k_{1,2}  |  \hat{H} e^{-i \hat{H} \theta} | \psi_{0} \rangle = e^{-i \lambda_1 \theta}  \lambda_1  (\vec{v}_{1,2}  + A \vec{w}_{1,2} ) 
\end{align}
and the included angle between vectors $ (\vec{v} +  \vec{w} )$ and $ (\vec{v} +  A \vec{w} )$ is $\tau_k  - \pi /2+ \arccos \{ \lambda_1/ |\lambda_1| \}$, because $\lambda_1$ may be negative. Now we have explicit expressions for
\begin{align}& c_{\psi,k} = \cos^2 \tau_k =  (1-\sin^2 \tau_k) = 1- \nonumber \\
& \!  \! \frac{\left(A R^2+(-1)^{k \! - \! 1}(A+1) \cos (\beta) R+1\right)^2}{\! \left(R^2 \! + \! 2 (-1)^{k \! - \! 1} \cos (\beta ) R \! + \! 1\right)    \!  \left(A^2 R^2 \! + \! 2 (-1)^{k \! - \! 1} A \cos (\beta) R \! + \! 1\right)}   , \label{coeffs}  \end{align} 
with \begin{equation}  \!  \!  \{ \!  R_{1}, \! R_{2} \} \!  \mapsto \! \left\{ \frac{ | \vec{w}_1|}{|\vec{v}_1|} ,  \!\frac{ | \vec{w}_2|}{|\vec{v}_2|} \right\} \! = \! \left\{ \tan(\alpha) \! \tan(\gamma),  \frac{\tan(\gamma)}{ \tan(\alpha)} \! \right\} , \! \end{equation} and  
$ \beta  =  \arccos \{ \vec{v_1}.\vec{w_1}/|\vec{v_1}||\vec{w_1}| \} =\chi - (\lambda_2 - \lambda_1) \theta  $. For angles $\{ \alpha, \beta, \gamma \}$ expectation value $\langle \psi_0 | \hat{H} \hat{\Phi}_{\theta}  \hat{H} | \psi_0  \rangle$ gives
\begin{widetext}
\begin{equation} \! \mathcal{J}(\alpha, \beta, \gamma)\! = \! \frac{- 4  \left(\lambda _1-\lambda
   _2\right){}^2 s^2[2 \alpha ] s^2[2 \gamma ] s^2\left[ \beta \right]}{\left(c[2 (\alpha -\gamma )]+c [2 (\alpha +\gamma )]+2 c \left[ \beta \right] s [2 \alpha ] s [2 \gamma ]-2\right) \! \left( c [2 (\alpha -\gamma )]+c [2 (\alpha +\gamma )]+2 c \left[ \beta \right] s [2 \alpha ] s [2 \gamma ]+2 \right)}
\end{equation}
\end{widetext}
writing sin as `$s$' and cos as `$c$'. This function is optimized by angles $\{ \alpha, \gamma \} \mapsto \pi / 4$, independent of the value of $\beta$ and giving a saturable bound: $ \langle \psi_0 | \hat{H} \hat{\Phi}_{\theta}  \hat{H} | \psi_0  \rangle \leq  \left(\lambda _1-\lambda_2\right){}^2 $. This is the upper bound on the Fisher information for any superposition of two eigenstates of the Hamiltonian. It is saturated by a probe state 
\begin{equation}
| \psi_{\text{opt}} \rangle =  (| \lambda_1 \rangle + e^{ i \chi } | \lambda_2 \rangle) / \sqrt{2} \; , 
\end{equation}
where $\chi \in [0,2 \pi)$, see FIG.\ref{sphere}. The result $\alpha \mapsto \pi/4 $ dictates an optimal measurement scheme with components:
\begin{equation}
| k_{\pm} \rangle =  (| \lambda_1 \rangle \pm | \lambda_2 \rangle) / \sqrt{2} \; ,  \label{best-measurement}
\end{equation}
also depicted in FIG.\ref{sphere}. All other basis elements $\in \{ | k \rangle \}$ span an orthogonal subspace. 

\emph{Generalization to Higher Dimensions:} It is clear that for a given Hamiltonian $\hat{H}$, the maximal Fisher information is then bounded from below  by $\left(\lambda _\uparrow - \lambda
   _\downarrow \right)^2 = || \hat{H} ||^2$ where $\lambda_\uparrow$ ($\lambda_\downarrow$) is the max (min) eigenvalue of $\hat{H}$, and $|| \hat{H}||=  \left(\lambda _\uparrow - \lambda_\downarrow \right)$ is the operator seminorm of the Hamiltonian \cite{MetrologyII}. A key property of the seminorm is that it gives an achievable upper bound to the variance: \begin{equation} || \hat{H} ||^2 \geq 4 \Delta^2 \hat{H}. \end{equation}  This allows a connection to be made between the qubit result with that for a fixed $| \psi_0 \rangle$ -- we saw earlier for a fixed input that $\mathcal{J} \leq 4 \Delta^2 \hat{H}$. Therefore the qubit maximum variance state must be the universally optimal state over the full Hilbert space, as it saturates its variance bound $\mathcal{J} =  || \hat{H} ||^2 $. The general result is
\begin{equation} \label{seminorm}
  \begin{array}{c}
    \text{max} \\ 
    | \psi_0 \rangle, \{ | k \rangle \} \\ 
  \end{array} \mathcal{J} =  || \hat{H} ||^2 \; .
\end{equation}
A corollary of Eq.\eqref{seminorm} is that no greater number of superposed energy eigenstates can improve on the Fisher information provided by the maximum variance state $(|\lambda_{\uparrow}  \rangle + e^{i \chi} | \lambda_{\downarrow} \rangle )/ \sqrt{2}$.

\emph{Photons:} In a  Mach-Zehnder interferometer (MZ) the estimated parameter $\theta$ is simply the phase difference between the two interferometer paths. Spin  $\hat{J}_y$ plays the role of Hamiltonian, and the Casimir operator $\hat{J}^2$ is equivalently $\hat{n}/2(\hat{n}/2+1)$ where $\hat{n}$ is the total photon number operator. (So $j = n/2$ for states of fixed $n$.)  Eigen-equations are $\hat{J}^2 | j,m \rangle_i = j(j+1)| j,m \rangle_i$ and  $\hat{J}_i | j,m \rangle_i = m | j,m \rangle_i$ for $i \in \{x,y,z \}$. The maximum variance state for the MZ is
\begin{equation}
|  \psi_{opt} \rangle = \frac{1}{\sqrt{2}} ( |j, +j  \rangle_y + e^{i \chi} | j, -j \rangle_y ) \; ,  \label{NOON}
\end{equation}
i.e. a NOON state \cite{Durkin-Dowling,NOON-both}, rotated by $\pi/2$ around the $x$ axis, with arbitray phase $\chi$. Eq.\eqref{best-measurement} indicates that the optimal measurement basis has two elements $| k_{\pm} \rangle = ( | j, + j \rangle_y \pm \exp(i \xi) |j , -j \rangle_y ) / \sqrt{2}$.  Interestingly, the optimal measurement basis is not unique (in this case) and maximal precision is also recovered by $\{ | k \rangle \}$ corresponding to the eigenbasis of $\hat{J}_z$, a measurement of photon number difference between the two interferometer modes \cite{Durkin-Dowling}. Therefore, for a lossless MZ the precision limit is saturated only by NOON probes of Eq.\eqref{NOON}, and this is achievable with simple projection measurements, i.e. without generalised measurements -- compare  \cite{Fisher-Optimal-measurements,Phase-POVM}. These optimal projection measurements may be performed by linear optics and photon counting alone. (For any photon number space of $n$ photons, of the full group of Unitary operations, SU$(n+1)$ , only the SU$(2)$ subset is needed.) The Cram\'{e}r-Rao  inequality in the context of interferometry thus gives a precision known as the `Heisenberg' limit,   \cite{Heisenberg-Limit} : 
$(\delta \theta_\text{H})^2 \geq1/ \mathcal{J}_{\text{max}} =  1/(\lambda_{\uparrow} - \lambda_{\downarrow})^2 =  1/ 4j^2 = 1 / n^2 $.

\emph{Phase States}
As an example of a fixed probe, take the phase state  \cite{phase-state,Phase-POVM} 
\begin{equation}\label{phase-state}
|\psi_{0} \rangle \mapsto |j, \zeta \rangle = \frac{1}{\sqrt{2j+1}} \sum_{m=-j}^{j}e^{i m \zeta } |j,m \rangle_y \; ,
\end{equation}
(parameterized by $\zeta \in \mathbbm{R}$). In a MZ the evolution is a translation of the parameter, $e^{-i \theta \hat{J}_y } |j, \zeta \rangle = |j, \zeta - \theta \rangle$. These states are of interest because it is expected that their precision scales $ \mathcal{J}\propto j^2$, close to the Heisenberg limit.  They have the peculiar feature that, for any $m$,
\begin{equation}
\arg \: _z \! \langle  j, m | \hat{O} | j, \theta \rangle = - j \pi/2 \;  , \; \forall \; \hat{O} \in SO(2j+1)  \; ,
\end{equation}
due to the properties \cite{Sakurai-Wigner} of the Wigner rotation elements $_z \langle j,  m_1 |  e^{-i \hat{J}_y \theta} | j, m_2 \rangle_{z} \in \mathbbm{R}$. It follows that choosing $| k \rangle = \hat{O} |j,m \rangle_z$ makes $ \phi_k$ independent of $\theta$, i.e. $\tau_k = \dot{\phi}_k= \mathcal{K} = 0$. Therefore Eq.\eqref{Fisher1} gives $\mathcal{J}_{\text{max}}(\theta) \mapsto 4 \langle \hat{J}^2_y \rangle = 4j(j+1)/3$, with the desired scaling.

\emph{Summary and Outlook}: The formalism developed here incorporates all aspects of quantum parameter estimation explicitly; probe $| \psi_{0} \rangle $, dynamics $\hat{H}$, and measurement, $\{ | k \rangle \}$, clarifying how precision is determined by the interplay of all three. We introduced the information complement $\mathcal{K}$, the minimization of which allowed the optimal measurement to be found for a fixed probe. The greatest precision was found in the qubit subspace of the maximal variance input state. This ultimate precision is  completely defined by the difference of the extremal energy eigenvalues  --  no deeper dynamical structure is relevant, nor is the dimension of the Hilbert space.  
 
Despite its many advantages (such as its easy extension to mixed states), if precision limits had been found using the quantum Fisher Information \cite{Metrology, MetrologyII}, nothing could be  learned about the relative performance of various measurement schemes, as that approach assumes an unknown optimal measurement. But for specific dynamical processes (e.g. photon interferometry) there exist real physical restrictions on the types of probe and measurement that may be employed.  For a restricted set that excludes an optimal measurement,  maximizing the classical Fisher information over the available measurements will allow the locally optimal probe to be recovered.

In future, it may prove fruitful to develop the Fisher Operator approach to incorporate evolution of mixed states, governed by completely positive maps and generalised measurements. 

This work was carried out under a contract with Mission Critical Technologies at NASA Ames Research Center. The author would like to thank Vadim Smelyanskiy and Gen Kimura for useful discussions.

\bibliographystyle{apsrev}

\end{document}